\documentstyle[preprint]{aastex}
\begin{document}

\title{Photometry and Spectroscopy of the Optical Companion to the Pulsar PSR
J1740-5340 in the Globular Cluster NGC 6397}

\author{J.~Kaluzny\altaffilmark{1}, S.~M. Rucinski\altaffilmark{2}, and
I.~B.~Thompson\altaffilmark{3}}

\altaffiltext{1}{Copernicus Astronomical Center, Bartycka 18,
00-716 Warsaw, Poland}
\altaffiltext{2}{Department of Astronomy, University of Toronto,
Toronto,
ON, Canada}
\altaffiltext{3}{Carnegie Observatories, 813 Santa Barbara St.,
Pasadena, CA 91101-1292}

\begin{abstract}

We present photometric and spectroscopic observations of the optical
companion to the millisecond radio pulsar PSR J1740-5340 in the globular
cluster NGC 6397. An analysis of the photometric variability in the
$B$-, $V$-, and $I$-bands indicates an inclination of the system of
43.9 $\pm$ 2.1 degrees if the optical companion fills its Roche lobe
(a semi-detached configuration). The spectroscopic data show a radial
velocity variation with a semi-amplitude of $K = 137.2 \pm 2.4$ km/sec,
and
a system velocity $\gamma$ = 17.6 $\pm$ 1.5 km/sec, consistent with
cluster
membership.  We use these results to derive a mass of the optical
companion of $M_{1}=0.296 \pm  0.034$~$M_{\sun}$ and
$M_{2}=1.53 \pm  0.19$~$M_{\sun}$ for the pulsar.
There is evidence for secular change of the amplitude
of the optical light curve of the variable measured over seven years.
The change does not have interpretation and its presence
complicates reliable determination of the  absolute
parameters of the binary.
\end{abstract}

\section{Introduction}

The millisecond radio pulsar  PSR J1740-5340 was discovered in the
field of the globular cluster NGC~6397  by D'Amico et al. (2001a) in the
course of a
survey conducted with the Parkes radio telescope.  Follow up pulse
timing
observations (D'Amico et al. 2001b) led to a determination of several
parameters of the system, including orbital period, projected
semi-major axis and mass function.  Ferraro et al. (2001)
identified the optical companion of the pulsar with a variable object
detected earlier by Taylor et al. (2001).  Photometry presented by both
groups shows that the optical companion to the MSP is a relatively
bright star, $V_{max}\approx 16.7$, located slightly to red of the
turn-off
region on the cluster color-magnitude diagram.  Remarkably, despite
being located only 26$\arcsec$ from the cluster center, the variable is
a relatively isolated star (see Fig. 2 in Ferraro et al. 2001),
permitting optical observations with ground-based telescopes. In this
paper we present the results of photometric and spectroscopic
observations obtained to determine the  masses for both components of
the binary.

The evolutionary status of PSR J1740-5340 has been discussed
by Burderi et al. (2002) and Ergma \& Sarna (2002). Both groups present
some
detailed scenarios which attempt to  explain the current status of the
binary and the observed position of its optical component on the
cluster color-magnitude diagram.
The system has been
detected in the X-ray domain with the $Chandra$ observatory by Grindlay
et al. (2001, 2002).

\section{Observations}

\subsection{Photometric Observations}

Observations in $B, V$, and $I$ filters were obtained with the 2K$^2$
TEK\#5 CCD camera on the 2.5-m du Pont telescope at Las Campanas
Observatory (LCO) in May and June 2002. The cluster was observed on 7
nights for a total of  32 hours.  The camera has a pixel scale of
0.259$\arcsec$/pixel.  Only 600 rows of the CCD were read out during
observations of the cluster in order to reduce dead time between
exposures, resulting in a field of view of 8.65x2.60 arcmin$^2$ centered
on the cluster.

Preliminary processing of the CCD frames was done with the
IRAF-CCDPROC package.\footnote{ IRAF is distributed by the National
Optical Astronomy Observatories, which are operated by the Association
of Universities for Research in Astronomy, Inc., under cooperative
agreement with the NSF.} For the photometric analysis we  used stacked
images each consisting  of 3 to 8 individual short exposures.
All co-added  images span less than 10 minutes and in most cases less
than 6 minutes.  In the order of $B$, $V$ and $I$-bands,
the total number of stacked images is  69,  197 and 59,
while average effective exposure times
are 136~s, 75~s and 50~s. In the same order of the bands,
the average seeing for the stacked images is 1.04\arcsec, 1.03\arcsec
and
0.87\arcsec.

We used the ISIS-2.1 image subtraction package (Alard \& Lupton 1998;
Alard 2000) to extract differential light curves of the optical
counterpart of the pulsar, following the prescription given in the
ISIS.V2.1 manual. For each band a template image was constructed from
several stacked frames of the best image quality.  Magnitude zero
points for the  ISIS differential light curves were measured from the
template images using the DAOPHOT/ALLSTAR software package (Stetson
1987), and aperture corrections were measured with the DAOGROW program
(Stetson 1990). Instrumental magnitudes were transformed to to the
standard $BVI$ system using the following relations:
   \begin{eqnarray}
V=v-0.0134(B-V)+0.1279(X-1.25)+0.7737\\
B=b-0.0690(B-V)+0.2243(X-1.25)+1.1254\\
I=i+0.0059(V-I)+0.0382(X-1.25)+1.3080
   \end{eqnarray}
where $b$, $v$ and $i$ are instrumental magnitudes and $X$ is the
air-mass.  The coefficients of the transformation were measured from 12
observations of 5 Landolt fields (Landolt 1992) obtained on the night
May 2 (UT).  The standard fields were observed over a range of air-mass
covering 1.16 to 2.00.  Figure 1 shows the residuals of the photometric
solution. We conclude that the
uncertainties of the zero points of our photometry do not exceed
0.02~mag.

$BVI$ light and color curves of the variable are presented in Figure 2
and Figure 3.  We phased our observations using the ephemeris
derived from radio timing observations from D'Amico et al. (2001b).
The  phase was shifted by 0.25 so that at phase 0.0 the optical
component is in conjunction (in front of the radio pulsar).\footnote{The
ephemeris
derived by D'Amico et al. (2001b) locates  phase 0.0  at the ascending
node of the pulsar orbit. This convention implies that for an
inclination $i=90\deg$ at phase 0.75, the observer sees the side of the
companion facing the pulsar. In our convention the observer sees that
side of the companion at phase 0.50.}
Throughout this paper we use the ephemeris:
   \begin{eqnarray}
HJD (Min~I) = 2\,451\,750.549336(7) + E \times 1.35405939(5)
   \end{eqnarray}
where the numbers in parentheses represent formal  $3\sigma$
uncertainties, in units of the last significant digit, 
of the very high precision determination of
D'Amico et al.
A program based on the KWEE algorithm (Kwee \& van Woerden 1956) was
used to determine moments of both  minima for the phased light curves.
It was found that after averaging the results for all 3 filters
the observed photometric minima occurred $0.006\pm 0.004$ of the orbital
period earlier than times of minima predicted by the above radio ephemeris.

The phase coverage is generally good with the exception of the
quadrature at phase 0.75 which is covered by only a few data points
in the $V$ filter and only single data points  in the  $B$ and $I$
filters.  The light curves show symmetric minima  and there is no
indication for any significant difference in the light level at the two
quadratures.  The color curves show some reddening near phase 0.5. All
of
these features indicate that the observed variability is caused mostly,
if not entirely, by ellipsoidal effects. In particular, the slight
difference in the depths of the minima (with the secondary minimum
being deeper) is consistent with such an interpretation.

In Table 1 we list $BVI$ magnitudes observed at the extrema of the
light curves as  determined by parabolic fits to points located with
phase $\pm 0.05$ from a given extrema. The full amplitude of the light
variation is 0.172(4), 0.152(4) and 0.139(5) magnitude
for the $B$, $V$ and $I$
bands, respectively.  We note that the HST observations collected in
April 1999 indicate a noticeably larger amplitude. Taylor et al. (2001)
report an amplitude of 0.21~mag in an unspecified band. However, the
$V$ band is the bluest of all bands they consider in their study.
The lower limit on the amplitude of variation in the $H\alpha$
band was 0.20 mag from photometry published by Ferraro et al. (2001).
We have extracted a $V$-band light curve of the
variable from time series observations of NGC~6397 obtained in July
1995  with the 0.9m telescope at the CTIO (Kaluzny 1997).  The data were
reduced using the image subtraction technique, and the derived light
curve
is presented in Figure 4. The scatter visible in the light curve near
phase 0.5 is consistent with the formal errors of the photometry.
The zero point of the 1995 photometry was set to the
zero point of the the 2002 data by comparing  photometry of several
stars located in the field of the variable. We
conclude that in 1995 the amplitude of variability in the $V$ band was
about 0.23 mag and  that the
variable showed the same level of maximum light in 1995 and 2002.
However, the system was fainter at secondary minimum  in 2002 in
comparison with the 1995 observations.  We comment on possible
interpretations of these changes in the last section of the paper.

\subsection{Spectroscopic Observations}

Spectra were obtained with the B\&C spectrograph on the 6.5-m Baade
telescope at LCO on the nights of
(UT dates) 30 May, 1 - 2 June, 8 June, and  29 July, 2002.
The spectral resolution was 2.0 A, and
the wavelength coverage was 3880 A to 5510 A (May and early June) and
3760 A to 5380 A (data taken on June 7 and July 29). The
weather conditions were
poor, with most of the data taken through clouds with seeing between
1.0 arcsec and 1.5 arcsec. The integration time for all exposures was
1800
seconds. Observations were also obtained of the velocity templates
HD 74000 (sdF6, velocity = +204.2 km/sec) in May and June, and HD 116064
(sdF0, velocity = +143.4 km/sec) in July.
These templates were selected from among low-metallicity stars to
match the program star spectra.

Figure 5 shows a mean spectrum of the optical companion of PSR
J1740-5340 from our observations on June 8 UT and July 29 UT, 2002, all
shifted to the systemic velocity of the binary (see below). We also
show for comparison a spectrum of the template star HD 74000. Despite
the position of the optical system well off the cluster main sequence,
the spectra of the stars are remarkably similar.   The metallicity of HD
74000 is  $\rm{[Fe/H] = -2.07}$  (Beveridge \& Sneden 1994) compared to
the cluster metallicity of [Fe/H] = -1.95 (Harris 1996).

Velocities of the optical companion were measured using the RVSAO
cross-correlation package (Kurtz and Mink 1998) running under IRAF, and
errors were adopted  from the XCSAO program. The results are
given in Table 2.   HD 74000 was used as the template for all of the
May and June observations, and HD 116064 for the July observations. The
relative velocity (HD 74000 $-$ HD 116064) was measured to be 56.2
$\pm$ 2.6 km/sec compared to our adopted value of 60.8 km/sec.

We made a least squares fit to the radial velocity data using GaussFit
(McArthur et al. 1994), solving for the system velocity $\gamma$ and
the velocity amplitude of the optical companion. We assumed a circular
orbit, and adopted the ephemeris given by Eq. (4). The results were
$K_1 = 136.03 \pm 2.75$ km/sec, and $\gamma = 17.62 \pm 1.69$ km/sec.
The heliocentric velocity of NGC 6397 is 19.2 km/sec (Gebhardt et al.
1995), and we conclude that this system is a cluster member. It has to
be noted, however, that due to the non-spherical shape of the optical
companion its radial velocity curve is expected to show some
departures from purely sinusoidal motion. Therefore the above values of
$K_1$ and $\gamma$ have to be treated as  first approximations  only
for further improvement in a combined photometric-spectroscopic
solution given in the next section.  The radial velocity observations
are presented in Figure 6 using the ephemeris from Equation 4.  Figure
6 also shows the GaussFit solution, and the velocity curve of the
pulsar itself ($K_2 = 26.6121\pm 0.0004$~km/s, as
derived from the radio pulsar timing data,
D'Amico et al 2001b)  shifted to a systemic velocity of 17.6 km/sec.

\section{Photometric and Spectroscopic Solutions}

To solve the  light and velocity curves of the optical component of
the binary pulsar we used the Wilson-Deviney (1971, hereafter W-D)
model as implemented in the 1986 version of the code. The code  is
described in some detail by Wilson (1979) and by Leung \& Wilson
(1977).  The MINGA\footnote{
The MINGA package can be obtained from
http://www.camk.edu.pl/$\sim$plewa}
 minimization package (Plewa 1988)
was used for the actual fitting of the observed  light curves and the
derivation of the system parameters. It was assumed that the only source
of light in wavelengths covered by our data is the surface of the
non-degenerate component. Using the $T_{eff}$ versus
$(B-V)$ calibration of Alonso et al. (1996) we estimated
an effective temperature $T_{1} = 5630$~K.
Appropriate values of the limb darkening coefficients were taken from
Wade \& Rucinski (1985): $X_{B}=0.679$, $X_{V}=0.577$,  $X_{I}=0.40$.
The gravity brightening coefficient was set to $g_{1}=0.32$, an
appropriate value for stars with convective envelopes.
The secondary component was assumed to be dark and very small so that it
does not influence the shape of the light curve. This was achieved by
setting $T_{2}=1000$~K and adopting a very high value of the
gravitational
potential $\Omega_{2}=900$.

There are some arguments for assuming that the binary is in
a semi-detached configuration, filling its Roche lobe. This case is
advocated by Burderi et al. (2002) who point out that a
relatively high rate of mass loss of
$\dot{M} \leq 10^{-10} M_{\odot}$~${\rm yr^{-1}}$ is  implied by
estimates of the mass of the gaseous envelope causing  the radio
eclipses
of PSR J1740-5340 to last about 40\% of the orbital period.
Further evidence supporting  a semidetached configuration is the fact
that
the binary harbors an extended  x-ray source, suggestive that the x-ray
flux results from the interaction of a relativistic wind with mass loss
from
the optical companion (Grindlay et al. 2001). We show below that there
is no evidence for heating of the optical companion by the MSP, and so
the
mass loss is likely to arise from Roche lobe overflow.
In this case the mass needed to power the extended  X-ray source could
be supplied directly through the inner Lagrangian point.

However we cannot completely reject the possibility that the binary has
a detached configuration, and in the following two sub-sections
we present light and radial velocity solutions
for both semi-detached and detached configurations.

\subsection{Semi-detached configuration}

The  shape of optical light curves of the binary
indicates that observed variability is dominated by
ellipsiodal effect. Experiments show that the light curves
can  be in fact quite well approximated by a sine function. That means
that
any model used to reproduce the observed light curves could have
no more than one free parameter, the sine-curve amplitude.
In our light curve solution conducted
for the assumed semi-detached configuration that free parameter
translates into $i$, the inclination of the binary orbit.

We started the analysis by modeling the light curves with an assumed
value of the mass ratio $q=m_{2}/m_{1}=5.11$, where the index $1$ refers
to
the optical component, and where  $q$ is  derived from the $K$
amplitudes
listed in the previous section. The W-D code was run in Mode-4, the
configuration with the primary component filling its Roche lobe.

The solution converged to a model with inclination $i=44.12\pm
2.15$~degrees.  The next step involved a determination of the radial
velocity amplitude $K_1$ by fitting the observed velocity curve with
model curves generated with the  W-D code. The W-D code allows one to
calculate the non-dimensional velocity curve, $v_{1th}$, and the $K_1$
amplitude is calculated  using the relation:
   \begin{eqnarray}
v_{1obs} = v_{1th}\times K_1 \times sin(i) \times(1+1/q) + \gamma
   \end{eqnarray}
where $v_{1obs}$ is the observed photocenter velocity and $\gamma$ is
the systemic velocity. We  derive  $K_1=137.2\pm 2.4$ km/s and
$\gamma=17.6\pm 1.5$ km/s. This in turn gives a new value of the mass
ratio $q=K_2/K_1=5.15$ which was used to get an improved solution
of the light curves. This procedure converged after one more iteration
and we derive  the following set of final parameters:
$i=43.9\pm 2.1$ degrees, $K_1=137.2\pm 2.4$ km/s and
$\gamma=17.6\pm 1.5$ km/s.

In Figure 7 we show the residuals for the final solution of the light
curves.  Note that while the fit is generally good, some systematic
errors  are present in fits obtained for the $V$ and $B$ bands  at the
level of 0.01 mag.  These systematic errors cannot be corrected without
introducing some additional features to the pure W-D model. The
deviations have a tendency to diminish with increasing wavelength and
indicate the presence of some extra light near phase 0.5 where we see
the hemisphere of the optical component facing the pulsar.  Some
irradiation processes may be responsible for this effect.  Figure 8
shows residuals for the model fit to the radial velocity curve.  The
observed residuals are consistent with the formal errors of the
measured velocities. Figure 9 shows how the $\chi^{2}$ statistic
changes as a  function of assumed inclination of the binary.  The fit
is excellent with a well defined minimum in $\chi^{2}$ at
$i\sim44$~deg.

Using these values of the radial velocity amplitudes and the
inclination we obtain masses of the components of the binary pulsar PSR
J1740-5340 of $M_1=0.296\pm 0.034~M_{\odot}$ and $M_2=1.53 \pm
0.19~M_{\odot}$.  Note that the  errors of the masses are dominated by
the accuracy of the derived inclination.  The absolute size of the
orbit is $A=6.30\pm 0.09 \, R_\odot$ and  the average radius of the
optical component is $R_1=1.67\pm 0.02 \, R_\odot $.

{\section{Detached configuration}

In the detached configuration, the light curve of the
system depends mostly on the inclination $i$ and the
gravitational potential $\Omega_{1}$. These two parameters are
strongly correlated and therefore we decided to obtain a grid
of solutions for several fixed values of $i$.
The results are given  in Table 3 which lists the adopted value of
inclination,
the average relative radius of the optical component,
its ratio  to the inner Roche lobe radius,
and the $\chi^{2}$ of the model.
Note that for a given mass ratio $q$, the  radius $r_{1}$ is
only a function of the gravitational potential $\Omega_{1}$.
Table 3 shows that fits of comparable quality were obtained for the
whole range of
adopted inclinations. In particular the solution  for $i=45$ degrees is
very close to the semi-detached configuration considered in the previous
section. Taken at face value, the fits for
the detached case and $i>~50$ are marginally better
that the fit derived  for the semi-detached case. Hence we believe
that we cannot with confidence determine  the configuration
of the binary  solely from the light curve solution.
Table 4 lists some absolute parameters of the binary
corresponding to the solutions from Table 3. Note that the formal
errors of the masses in Table~4 do not include an uncertainty
in the inclination as it was fixed for each entry in the table.

An additional and entirely independent constraint on the system
inclination arises from  information about its cluster membership.
Reid (1998) and Reid \& Gizis (1998) have used  main-sequence
fitting  to derive  $V$-band apparent distance
moduli for NGC 6397 of $12.80\pm 0.1$ and $12.69\pm 0.15$, respectively.
Adopting a
distance modulus $(m-M)_{\rm V}=12.72\pm 0.18$ and using $<V>=16.71$
(see Table 1) we obtain $<M_{\rm V}>=3.99\pm 0.18$
as an estimate of the average absolute
magnitude of the optical component
for an assumed reddening of $E(B-V)$ = 0.18 (Reid \& Giziz 1998).
This gives $<M_{bol}>=3.72\pm 0.18$ for
$BC=-0.27$
which is appropriate for the observed color and metallicity of the star
(Houdashelt et al. 2000). Using the relation,
$R/R_{\odot} = (L/L_{\odot})^{1/2}\times(T_{eff\odot}/T_{eff})^{2}$,
we obtain $R_{1}=1.68\pm 0.14 \, R_{\odot}$. This apparently
eliminates all solutions with $i > 47$ degrees listed in Table 4.

A further constraint on the inclination could come from a measurement
of the rotational velocity of the optical companion. Our spectra do not
have high enough resolution, however the optical companion is bright
enough that such measurements could be made with echelle spectrographs
on 6.5-m class telescopes.

\section{Discussion}

An additional complication to the  interpretation of the optical
observations of the optical companion is the fact that its light curve
clearly evolves on a time scale of a few years.  The range of
variability in the $V$ band has changed from $\Delta V\approx 0.23$ mag
in 1995 to $\Delta V\approx 0.15$ in 2002.  We discuss briefly 3
possible interpretations of these changes.

1. {\it Secular change of the inclination of the orbit of binary}:
Assuming a  semi-detached configuration the observed change of $\Delta
V$  requires a change of orbital inclination of about  8 degrees (from
$i=52$~deg in 1995 to $i=44$~deg in 2002).  There are a few eclipsing
binaries with observed variation in the inclination of their orbital
plane due to an interaction with a third body  (Drechsel et al. 1994;
Milone et al. 2000).  However the rate of variation is at a level of a
few tenths of a degree per year at best.  Pulsar timing observations
covering a 6 month interval (D'Amico et al. 2001b) show no evidence for
a dynamical interaction of PSR J1740-5340 with a hypothetical "third
body".  Further radio timing  observations of the pulsar should provide
very strong limits on any changes of orientation of the orbital plane
of the binary.

2. {\it Variable "third light" contributions to the light curve}:
Luminous streams of gas around pulsar or variable
heating of the optical component of the system could lead
to changes in the light curve. However in this  case  we should observe
not only changes in the  amplitude of the light curves but also changes
in the maximum observed light. One may estimate how much
of 3rd light is needed to diminish $\Delta V$ from $\approx 0.23$
(1995 season) to $\approx 0.15$ (2002 season),
 where $\Delta V$ is magnitude difference
for phases 0.25 and 0.50. Denoting the flux level at phase 0.25 in 1995
season
by $l_{1V}$ we obtain $l_{3V}=0.48 \, l_{1V}$. That in turn implies that
at quadrature the system would be brighter by 0.42 mag in 2002 season
as compared with 1995 season.  Our data show no indications for
any change of light level at quadratures between the 1995 and 2002
seasons, which would exceed 0.02--0.03 mag.}

3. {\it Intrinsic variability of the optical companion}: It is possible
that the observed variability is intrinsic to the optical companion, 
perhaps in
the form of star-spots. The companion has a high rotational velocity if
it is tidally locked ($v_{rot} \simeq 50$ km/sec)
and such a high rotational velocity in a fully convective atmosphere
normally leads to the formation of star-spots.
We have used WD code to perform some light curve simulations
for a model including one dark spot. The starting point was the light
curve
solution obtained for the semi-detached configuration and the 2002 data.
It turns out that one may indeed increase the depth of the
minimum observed at phase 0.5 to 0.23~mag, as seen in 1995 season,
by putting a dark spot in a region around inner Lagrangian point $L1$.
Specific parameters of such a  spot are $\Delta T=1000$K and radius
equal to 20~degrees (as seen from the center of the star).
The corresponding change of $V$ magnitude at quadratures would then
amount to only 0.02~mag.
It is worth to note in that context that light curves presented
by Ferraro et al. (2001; Fig. 4) show a clear asymmetry at
minimum light which occurs at phase 0.5. Spot hypothesis
offers a possible way to explain such a distortion.

We do not have a definite  explanation for the
observed variations in the amplitude of the light curves.
The hypothesis invoking stellar spots seems to be a viable
option for a moment.
Clearly further monitoring of the system would be desirable.

\section{Summary}

Our analysis indicates that the observed
modulation of the optical light  from the optical companion
of the  pulsar PSR J1740-5340 can be fully explained
by ellipsoidal variations. There is no indication for
any detectable light due to heating of the companion by radiation
from  the pulsar. However, the light curve amplitude appears to
evolve on a time scale of a few years, a phenomenon without
a clear interpretation which may systematically affect our
results. 

An analysis of the radial velocity curve  indicates that,
if the velocities are modeled with a simple sinusoidal variation,
neglecting the non-spherical shape of the optical component,
$K_1 = 137.2 \pm 2.4$ km/sec, and $\gamma = 17.6 \pm 1.5$ km/sec.
The measured systemic velocity of the system is
is consistent with cluster membership of the binary.
Gebhardt et al. (1995) obtained for NGC~6397 $\gamma = 19.2 \pm 0.5$
and measured $\Delta V_{rad} \approx 5.0$ km/sec at radius $r=30\arcsec$
from the cluster center. This observation is interesting in light of
speculations about the possible formation of the system in
a relatively recent three body interaction (eg. Grindlay et al. 2002).

The low amplitude of the observed light variations
and lack of optical eclipses precludes a
determination of the inclination of the orbit of the binary without
making any assumptions. A well constrained solution for the  light and
velocity data (taking into account the photocenter velocity correction)
was obtained by assuming a semi-detached configuration for the
binary, resulting in  $i=43.9\pm 2.1$ degrees for the system inclination and
$M_1=0.296\pm 0.034 \, M_{\odot}$ and $M_2=1.53\pm 0.19 \, M_{\odot}$
for the masses of optical companion and the pulsar, respectively.
Measurements of the masses of radio pulsars are know to show
a remarkably narrow Gaussian mass distribution
with $M=1.35\pm 0.04 \, M_{\odot}$ (Thorsett and Chakrabarty 1999),
and our measurement
of the mass of PSR J1740-5340 is consistent with this result.

Relaxation of the assumption of a semi-detached configuration leads to
light curve fits of similar quality  
for a wide range of inclinations, $44<i<90$~degrees.
If we assume a bolometric correction of the optical
companion of $BC=-0.27$ then the distance modulus of the cluster
constrains the inclination to a range of $44<i<47$ degrees.
In such a case our spectroscopic data give
$M_2 \ge 1.31 \pm 0.06 M_{\odot}$ as a lower limit on the pulsar mass.
The lower limit to the mass of the optical companion is
$M_2 \ge 0.255 \pm 0.007 M_{\odot}$.
The low mass of the optical companion of  PSR J1740-5340
resulting from our analysis supports evolutionary scenarios
presented recently by Burderi et al. (2002) and Ergma \& Sarna (2002).

\acknowledgments
JK was supported by the Polish KBN grant 5P03D004.21 and by
NSF grant AST-9819787. IT was supported by NSF grant AST-9819786.
SMR was supported by a grant from the Natural Sciences and Engineering
Research Council of Canada. We thank Jennifer Johnson and Steve Shectman
for sharing some  telescope time with us.
We are grateful to Bohdan Paczynski for useful comments on the
manuscript.

\clearpage

\begin{table}[ht]
\caption[]{\sc BVI magnitudes at extrema of light curves\\}
\begin{flushleft}
\begin{tabular}{lccc}
\hline\hline
 Phase & B        & V       & I \\
\hline
0.00 & 17.504(3) & 16.781(2)& 15.832(3)\\
0.25 & 17.357(3) & 16.647(3)& 15.706(3)\\
0.50 & 17.529(3) & 16.799(2)& 15.845(4)\\
0.75 & ~         & 16.647(6)&  ~       \\
\hline
\end{tabular}
~\\
Note: Internal errors are given in parentheses.
\end{flushleft}
\label{tab:phot1}
\end{table}

\clearpage

\begin{table}[ht]
\caption[]{\sc Radial velocities of the optical companion of
PSR~J1740-5340\\}
\begin{flushleft}
\begin{tabular}{lcrr}
\hline\hline
HJD \tablenotemark{a} & Phase & V[km/s]  & $\sigma$K \\
\hline
2424.579 & 0.785 & -105.35 &  17.04     \\
2424.598 & 0.798 & -110.31 &  17.77     \\
2424.625 & 0.818 & -110.80 &   9.11     \\
2424.652 & 0.838 &  -84.19 &   8.17     \\
2424.676 & 0.856 &  -76.78 &   6.78     \\
2424.704 & 0.876 &  -81.83 &   7.41     \\
2424.729 & 0.895 &  -67.08 &   6.67     \\
2424.763 & 0.920 &  -60.64 &   7.38     \\
2424.787 & 0.938 &  -53.80 &   9.18     \\
2424.815 & 0.959 &  -36.47 &  12.08     \\
2424.839 & 0.977 &  -21.75 &  15.59     \\
2424.877 & 0.004 &    7.81 &  16.79     \\
2425.591 & 0.531 &   -3.34 &  11.43     \\
2425.615 & 0.550 &  -30.00 &  18.01     \\
2425.878 & 0.744 & -115.51 &  11.12     \\
2426.567 & 0.244 &  151.80 &  11.17     \\
2426.591 & 0.270 &  160.33 &   9.35     \\
2426.615 & 0.288 &  157.36 &  10.92     \\
2426.713 & 0.358 &  121.86 &  12.63     \\
2426.736 & 0.388 &  117.63 &  14.28     \\
2428.785 & 0.885 &  -78.08 &   7.40     \\
2433.609 & 0.454 &   62.58 &   9.38     \\
2433.858 & 0.637 &  -83.88 &  11.63     \\
2484.499 & 0.032 &   50.19 &   4.63  \\
2484.523 & 0.050 &   70.68 &   4.35  \\
2484.553 & 0.072 &   73.76 &   5.62  \\
2484.576 & 0.089 &   79.94 &   5.63  \\
2484.600 & 0.107 &   90.81 &   7.44  \\
2484.623 & 0.124 &  107.70 &   8.52  \\
2484.654 & 0.147 &  130.84 &   8.62  \\
2484.678 & 0.164 &  130.32 &   9.18  \\
\tablenotetext{a}{HJD - 2450000.0}
\end{tabular}
~\\
\end{flushleft}
\label{tab:phot2}
\end{table}

\clearpage

\begin{table}[ht]
\caption[]{\sc Light curve solutions for detached
configuration and fixed inclination\\}
\begin{flushleft}
\begin{tabular}{lrlcc}
\hline\hline
$i[deg]$ & $\Omega_{1}$ & $<r_{1}>$ & $<r_{1}>/<r_{1in}>$ & $\chi^{2}$
\\
\hline
90 & 10.113(18) & 0.201(7)  & 0.75 & 285 \\
85 & 10.102(18) & 0.210(7)  & 0.79 & 281 \\
80 & 10.067(18) & 0.212(8)  & 0.80 & 271 \\
75 & 10.013(18) & 0.215(8)  & 0.81 & 262 \\
70 & 9.941(19)  & 0.219(8)  & 0.82 & 258 \\
65 & 9.850(19)  & 0.224(8)  & 0.84 & 258 \\
60 & 9.743(18)  & 0.230(9)  & 0.86 & 266 \\
55 & 9.620(18)  & 0.239(10) & 0.90 & 286 \\
50 & 9.489(17)  & 0.249(10) & 0.94 & 317 \\
47 & 9.412(10)  & 0.257(5)  & 0.96 & 336 \\
46 & 9.388(13)  & 0.259(7)  & 0.97 & 342 \\
45 & 9.365(11)  & 0.263(7)  & 0.99 & 345 \\
\end{tabular}
~\\
\end{flushleft}
\label{tab:sol1}
\end{table}

\clearpage


\begin{table}[ht]
\caption[]{\sc Absolute parameters corresponding to solutions from Table
3}
\begin{flushleft}
\begin{tabular}{lllccc}
\hline\hline
$i[deg]$ & $K_1$ & $A/R_{\odot}$& $<R_1>/R_{\odot}$ &
$M_1/M_{\odot}$&$M_2/M_{\odot}$\\
\hline
90 & 136.2 &4.38(6)  &0.88(3)  &0.099(3) & 0.51(2) \\
85 & 136.2 &4.40(6)  &0.92(3)  &0.100(3) & 0.51(2) \\
80 & 136.2 &4.45(7)  &0.94(4)  &0.104(3) & 0.53(2) \\
75 & 136.2 &4.53(7)  &0.98(4)  &0.110(3) & 0.56(3) \\
70 & 136.3 &4.66(7)  &1.02(4)  &0.120(4) & 0.61(3) \\
65 & 136.3 &4.83(7)  &1.08(4)  &0.133(4) & 0.68(3) \\
60 & 136.4 &5.06(7)  &1.16(5)  &0.153(4) & 0.78(4) \\
55 & 136.5 &5.35(8)  &1.28(5)  &0.181(5) & 0.93(4) \\
50 & 136.7 &5.72(8)  &1.42(6)  &0.222(7) & 1.14(4) \\
47 & 136.8 &5.99(9)  &1.54(2)  &0.255(7) & 1.31(6) \\
46 & 136.9 &6.09(9)  &1.58(4)  &0.268(8) & 1.38(6) \\
45 & 137.0 &6.20(9)  &1.63(4)  &0.282(8) & 1.46(7) \\
\end{tabular}
~\\
\end{flushleft}
\label{tab:sol2}
\end{table}

\clearpage

\begin{figure}
\figurenum{1}
\plotone{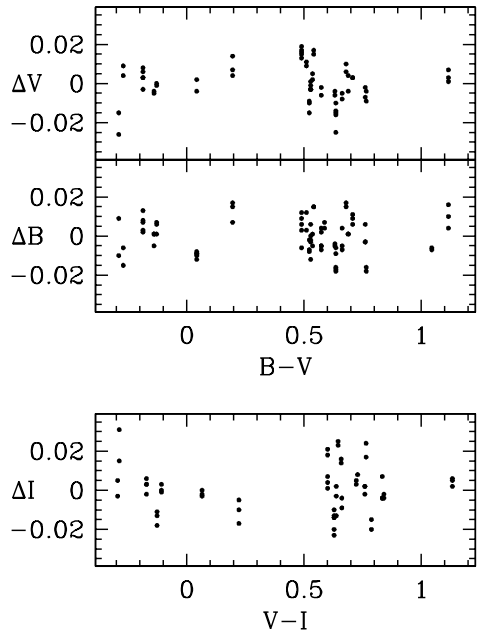}
\caption{The $BVI$ residuals of Landolt standard stars as a function
of color resulting from transformations defined by Eqs. (1)-(3).}
\end{figure}

\clearpage

\begin{figure}
\figurenum{2}
\plotone{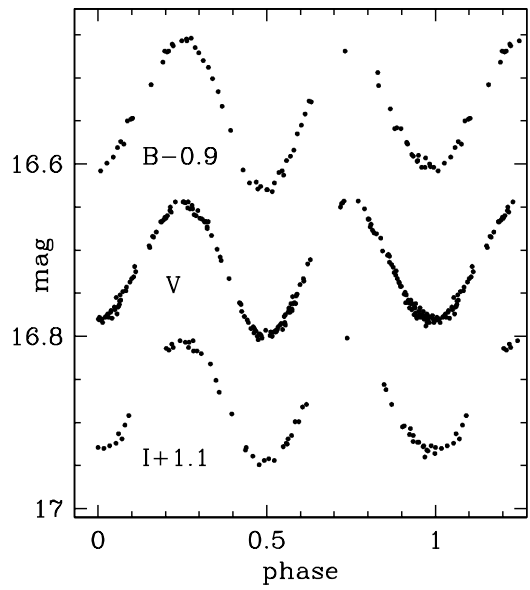}
\caption{$BVI$ light curves of the optical companion of the MSP PSR
J1740-5340 obtained in May-June 2002. Note that the data for the $B-$
and $I$-bands have been  shifted by $-0.9$ and $+1.1$ mag,
respectively }
\end{figure}

\clearpage

\begin{figure}
\figurenum{3}
\plotone{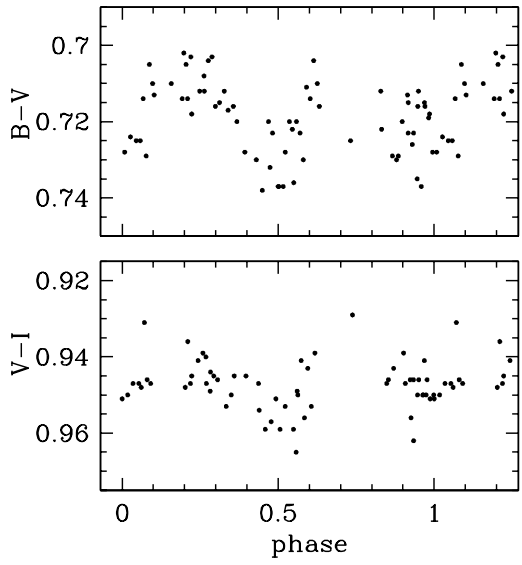}
\caption{ Color curves of the optical companion of
the MSP PSR J1740-5340.}
\end{figure}

\clearpage
\begin{figure}
\figurenum{4}
\plotone{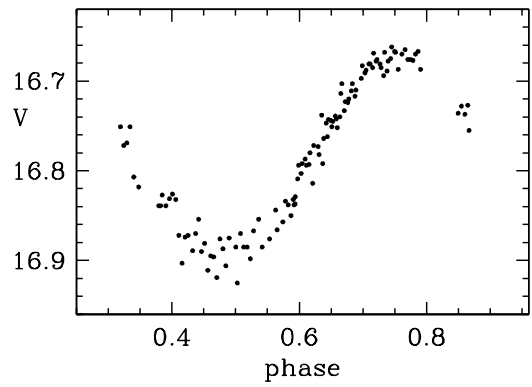}
\caption{V-band light curve of the optical companion of
the MSP PSR J1740-5340 obtained in July 1995.}
\end{figure}

\clearpage

\begin{figure}
\figurenum{5}
\plotone{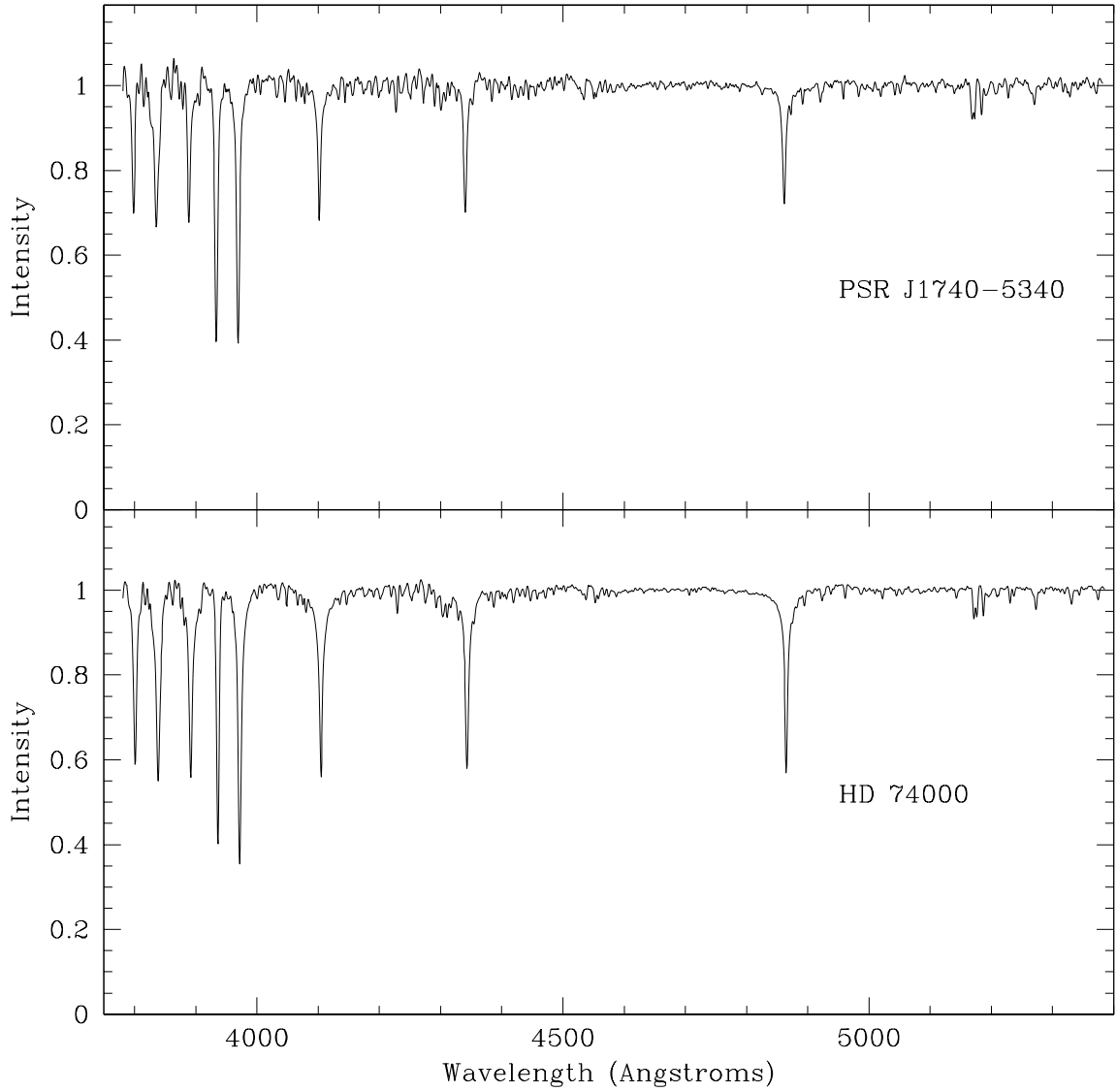}
\caption{The mean intensity-normalized
spectrum of the optical companion to PSR J1740-5340 (top)
together with a spectrum of the velocity template HD 74000, spectral
type
sdF6 (bottom).}
\end{figure}

\clearpage

\begin{figure}
\figurenum{6}
\plotone{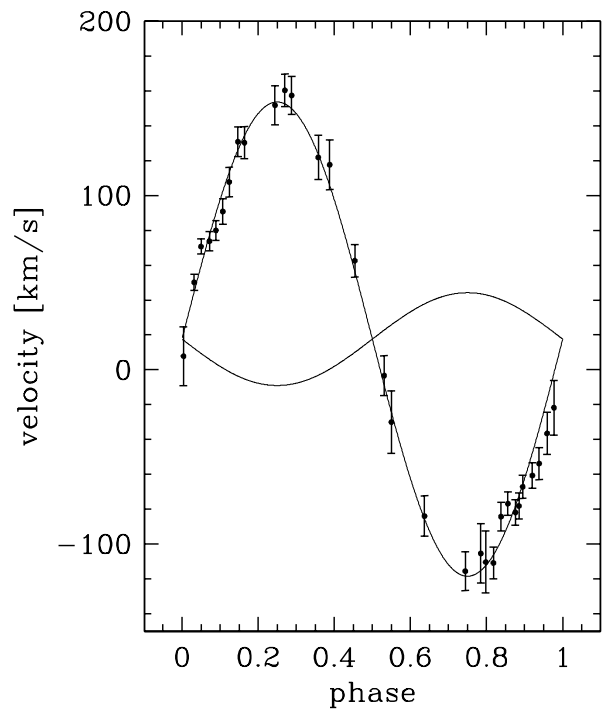}
\caption{The radial velocity measurements plotted with the circular
orbital solution. The low amplitude curve represents the orbital motion
of the pulsar as derived from timing observations.}
\end{figure}

\clearpage

\begin{figure}
\figurenum{7}
\plotone{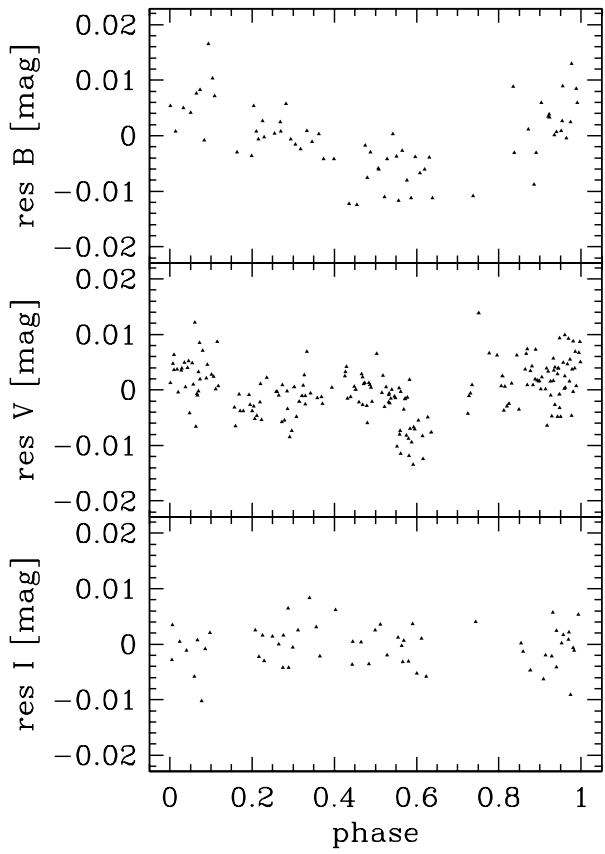}
\caption{
$O-C$ residuals  for the photometric solutions of the $BVI$ light
curves.
}
\end{figure}

\clearpage

\begin{figure}
\figurenum{8}
\plotone{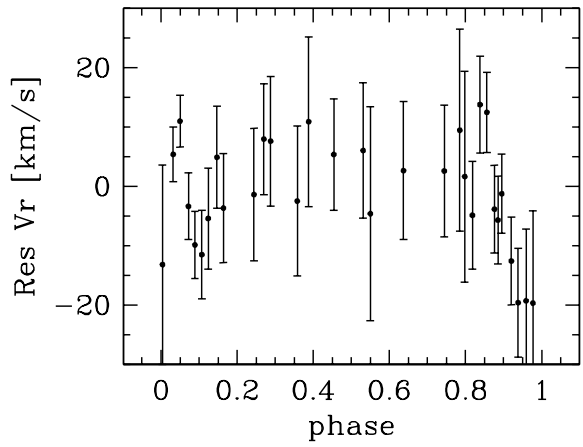}
\caption{The radial velocity residuals (observed minus calculated)
for the model corresponding to semi-detached configuration}.
\end{figure}
\clearpage

\begin{figure}
\figurenum{9}
\plotone{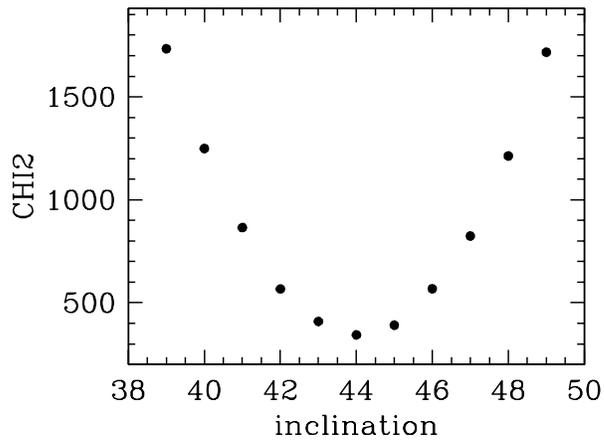}
\caption{$\chi^{2}$ statistic for light curve solution
as a function of assumed inclination.}
\end{figure}

\end{document}